\documentclass[aps,prl,twocolumn,longbibliography,nobibnotes]{revtex4-1}

\usepackage{natbib}
\usepackage{xspace}

\usepackage[sumlimits]{amsmath}
\usepackage{amsfonts,amssymb,amsthm}
\usepackage{graphicx}
\usepackage{physics}
\usepackage[margin=1in,footskip=0.25in]{geometry}

\usepackage{xr}
\usepackage{xcolor}

\newcommand{\TJWat}{IBM Quantum, IBM T.J. Watson Research Center, Yorktown Heights, NY 10598, USA}

\newcommand{\figfolder}[1]{}

\begin{document}

\title{Demonstration of a High-Fidelity CNOT for Fixed-Frequency Transmons with Engineered ZZ Suppression}

\author{A.~Kandala}
\email{akandala@us.ibm.com}
\author{K.~X.~Wei}
\email{xkwei@ibm.com}
\author{S.~Srinivasan}
\email{srikants@us.ibm.com}
\author{E.~Magesan}
\author{S.~Carnevale}
\author{G.~A.~Keefe}
\author{D.~Klaus}
\author{O.~Dial}
\author{D.~C.~McKay}
\email{dcmckay@us.ibm.com}

\affiliation{\TJWat}
\date{\today}

\begin{abstract}
Improving two-qubit gate performance and suppressing crosstalk are major, but often competing, challenges to achieving scalable quantum computation. In particular, increasing the coupling to realize faster gates has been intrinsically linked to enhanced crosstalk due to unwanted two-qubit terms in the Hamiltonian. Here, we demonstrate a novel coupling architecture for transmon qubits that circumvents the standard relationship between desired and undesired interaction rates. Using two fixed frequency coupling elements to tune the dressed level spacings, we demonstrate an intrinsic suppression of the static $ZZ$, while maintaining large effective coupling rates. Our architecture reveals no observable degradation of qubit coherence ($T_1,T_2 > 100~\mu s$) and, over a factor of 6 improvement in the ratio of desired to undesired coupling. Using the cross-resonance interaction we demonstrate a 180~ns single-pulse CNOT gate, and measure a CNOT fidelity of 99.77(2)$\%$ from interleaved randomized benchmarking.
\end{abstract}
\pacs{}

\maketitle

Quantum computing requires well-controlled, multi-qubit devices that offer speedup in certain tasks compared to their classical counterparts. Recently, there has been an explosion in device scaling, mostly based on superconducting qubits~\cite{huang:2020,arute:2019}. However, multi-qubit circuit fidelity, and ultimately the path to a fully fault tolerant architecture, is impeded by the tradeoff between crosstalk and gate speed. This tradeoff is implicit in the canonical cQED Hamiltonian for two transmons with fixed coupling($i=\{0,1\}$),
\begin{eqnarray}
H & = & \sum_{i=\{0,1\}} \left(\omega_i \hat{a}_{i}^{\dagger}\hat{a}_{i} + \frac{\alpha_i}{2}\hat{a}_{i}^{\dagger}\hat{a}_{i} \left[\hat{a}_{i}^{\dagger}\hat{a}_{i}-1\right]\right) + \nonumber \\
&& J (\hat{a}_0^{\dagger}+\hat{a}_0)(\hat{a}_1^{\dagger}+\hat{a}_1),  \label{eqn:mainh2}
\end{eqnarray} 
with frequencies $\omega_i$, anharmonicities $\alpha_i$ and coupling strength $J$ that can be engineered by a common bus resonator~\cite{dicarlo:2009} or direct capacitance~\cite{barends:2014}. The entanglement rate is set by $J$ for a number of two-qubit gates ~\cite{dicarlo:2009,poletto:2012,paraoanu:2006,chow:2013b,krinner:2020}, and so, a large $J$ is desirable for fast two-qubit entangling gates. This maximizes gate fidelity given finite qubit coherence. However, in this Hamiltonian, the dressed energy levels have a two-qubit frequency shift (to second order in $J$)~\cite{Magesan2020}
\begin{eqnarray}
ZZ & = & \omega_{11}-\omega_{01}-\omega_{10}+\omega_{00}, \\
& = & 2J^2 \frac{\alpha_0+\alpha_1}{(\Delta+\alpha_0)(\Delta-\alpha_1)}, \label{eqn:zzpert}
\end{eqnarray} 
where $\Delta$ is the qubit-qubit detuning. For fixed couplings, this interaction is an always-on source of error and is referred to as the static $ZZ$. It limits multi-qubit circuit performance~\cite{sundaresan:2020,jurcevic:2020,mckay:2019,krinner:2020b,wei:2020}, and is an impediment for realizing quantum error detection~\cite{takita:2017,andersen:2020}. The unfavorable quadratic scaling of the $ZZ$ error term puts a strict upper limit on $J$ in single coupler designs, leading to slow gates.

An alternative approach to mitigating crosstalk employs tunable coupling elements with large on/off ratios for $J$~\cite{ploeg:2007,chen:2014,weber:2017}. The introduction of tunable elements typically leads to additional decoherence and control complexity. More recent approaches have directly focused on suppressing the static $ZZ$ interaction by engineering the two-qubit level spacings. As seen from Eqn.~\ref{eqn:zzpert}, this can be achieved by coupling qubits with opposite signs of anharmonicity~\cite{zhao:2020,ku2020}. This effect can also be achieved by employing multiple coupling paths~\cite{mundada:2019,yan:2018,collodo:2020,xu:2020,sung:2020} with tunable elements. In both approaches, the suppression of static $ZZ$ results in clear improvements to simultaneous single qubit gate performance.

\begin{figure}[]
\includegraphics[width = \columnwidth]{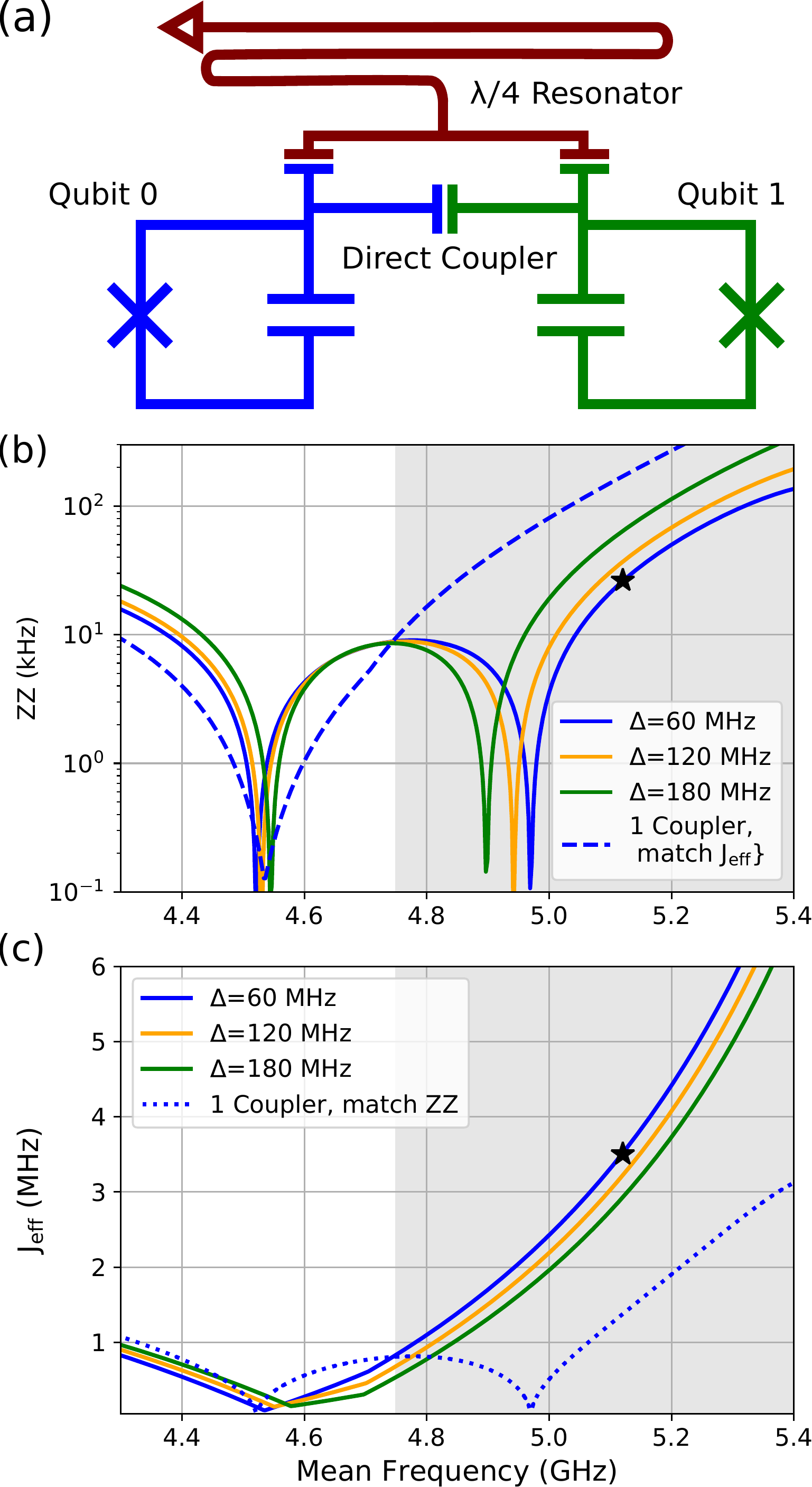}
\caption{(a) A circuit schematic of the device described in the main text (device A). The device consists of two fixed-frequency transmon qubits with a direct coupler and a $\lambda/4$ resonator (see parameters in the main text).  (b) For the values of  $g_1, g_2, J_0$ extracted from fitting the Hamiltonian model to device A, we look at $ZZ$ vs the mean qubit frequency at different qubit-qubit detunings. The experimental data for device A is highlighted by the star. The dashed line is the $ZZ$ for a pair of qubits with $\Delta=60$~MHz coupled via a single path (for example, a direct coupler) with the same effective $J$ as device A (see part (c)). (c) The effective $J$ for device A at different qubit-qubit detunings, the experiment value is the star. The dotted line is the effective $J$ for a $\Delta=60$~MHz direct coupler device with the same $ZZ$ as device A (see part (b)). The shaded region represents the frequency region where the multi-path coupler shows an improvement in $J_{\textrm{eff}}$/ZZ. \label{fig:schematic}}
\end{figure}

In this work, we demonstrate ZZ suppression by using multiple paths made purely from fixed-frequency, non-tunable elements. The lack of tunability means the circuit is simple to control and insensitive to noise. Nonetheless, it is shown to be robust to variations in circuit parameters such as the qubit frequencies. The result is a device with an effective $J$ of 3.5~MHz, yet a $ZZ$ of only 26~kHz. We explore the physics of the cross-resonance (CR) interaction~\cite{paraoanu:2006,rigetti:2010,chow:2011,sheldon:2016,magesan:2020,sundaresan:2020,ku2020} with this novel device architecture, and demonstrate a CNOT gate with a fidelity of 99.77(2)\%. To understand this device, we start with the Hamiltonian for two transmon qubits with multiple coupling paths,
\begin{eqnarray}
H & = & \sum_{i=\{0,1\}} \left(\omega_i \hat{a}_{i}^{\dagger}\hat{a}_{i} + \frac{\alpha_i}{2}\hat{a}_{i}^{\dagger}\hat{a}_{i} \left[\hat{a}_{i}^{\dagger}\hat{a}_{i}-1\right]\right)  \nonumber \\
&& +  J_0 (\hat{a}_0^{\dagger}+\hat{a}_0)(\hat{a}_1^{\dagger}+\hat{a}_1) + \sum_{j=0}^{N_{\textrm{bus}}} \omega_j \hat{b}_{j}^{\dagger}\hat{b}_{j} \nonumber \\
&&  +  \sum_{i=\{0,1\}} \sum_{j=0}^{N_{\textrm{bus}}} g_{i,j}(\hat{a}_i^{\dagger}+\hat{a}_i)(\hat{b}_j^{\dagger}+\hat{b}_j), \label{eqn:mainh}
\end{eqnarray} 
where $J_0$ is the direct exchange coupling, and $g_{i,j}$ is the coupling from qubit $i$ to harmonic resonator mode $j$. With coupling amplitudes $g_{i,j}, J_0$ of the appropriate sign, diagonalizing the Hamiltonian of Eqn.~\ref{eqn:mainh} results in contributions to the energy level shifts from the multiple coupling terms and leads to an effective cancellation of the static $ZZ$ interaction. Specifically, we show that for fairly accessible coupling amplitudes, the static $ZZ$ can be suppressed over a large range of qubit frequencies in the straddling regime $|\Delta| < |\alpha|$ (see Supplementary Information). In this work, we realize such a device Hamiltonian by simultaneously coupling two qubits with a $\lambda/4$ CPW resonator with its fundamental mode above both qubit frequencies and a direct capacitive coupler (short CPW section); for this geometry, $g_1, g_2, J_0 > 0$. A schematic is shown in Fig.~\ref{fig:schematic}(a). 

The central device discussed in this work, device A, has qubit frequencies $f_0 (f_1)$ = 5.1518 (5.0892)~GHz, and $\alpha_0 (\alpha_1)$ = -302 (-302)~MHz. During the course of this work, the average coherence properties for the qubits [Q0,Q1] were $T_1=[115(11),117(17)]~\mu s$ and $T_2=[129(14),139(32)]~\mu s$. The $\lambda/4$  resonator frequency is $f_{\textrm{bus}}$= 5.9638 GHz with $g_0 (g_1) = 88.5(87.5)$ MHz estimated from photon number splitting. For these numbers, we see that exclusion of higher bus modes (next mode at $\sim$ 18 GHz) does not significantly affect our analysis of the device Hamiltonian. We measure a static $ZZ$ of 26 kHz, and use that to fit to a $J_0 =$ 6.2 MHz in Eqn.~\ref{eqn:mainh}. The $ZZ$ cancellation, and the CR gate speed are both dependent on the qubit frequencies and so we plot them as a function of mean qubit frequency and for different qubit detunings, in Fig.~\ref{fig:schematic} (b) and (c), respectively. Here, the CR gate speed is quantified in terms of an effective $J$ ($J_{\textrm{eff}}$). We numerically calculate $\mu=ZX/\Omega$ ($\Omega$ is the CR drive power) for the multi-path coupler and define 
\begin{equation}
J_{\textrm{eff}} = \mu \frac{(\alpha+\Delta)\Delta}{\alpha},
\end{equation}
i.e., the value of $J$ from Eqn.~\ref{eqn:mainh2} for a single coupler that would provide the same $\mu$~\cite{magesan:2020}. Fig.~\ref{fig:schematic} (b) displays two points of sign changes of the static $ZZ$ through zero. The $ZZ=0$ point at the lower mean frequency trivially corresponds to $J_{\textrm{eff}} \sim 0$ as seen in Fig.~\ref{fig:schematic} (c). However, crucially for CR operation, the second $ZZ=0$ point at higher mean frequency maintains a finite $J_{\textrm{eff}}$. For the  $J_{\textrm{eff}}$ measured on device A, the static $ZZ$ arising from an equivalent standard direct coupler is also shown for comparison in Fig.~\ref{fig:schematic} (b). The difference between the dashed and solid lines demonstrate that the multi-path couplers break the typical fixed relationship between $J$ and $ZZ$ set by Eqn.~\ref{eqn:zzpert}. This manifests as a significant increase in the ratio of the desired coupling to the undesired coupling, $J_{\textrm{eff}}/ZZ$, over a broad range of qubit frequencies, despite the narrow bandwidth of the zero in $ZZ$ and without sacrificing the strength of $J$. Such a range of qubit frequencies is indicated by the the shaded region in Fig.~\ref{fig:schematic} (b),(c). Indeed, while device A does not operate at a $ZZ=0$ point, we measure $J_{\textrm{eff}}/ZZ \approx 130$ with $J_{\textrm{eff}}=3.5$~MHz in contrast to an equivalent-$J$ single coupler, where the ratio, at the same $J$ and $\Delta$, is $\approx$ 20.  In practice, there is a limit to how far the qubit frequencies should be above the $ZZ=0$ point set by the desired absolute value of the $ZZ$, which will increase idle and simultaneous single qubit gate error.

\begin{figure}[]
\includegraphics[width = \columnwidth]{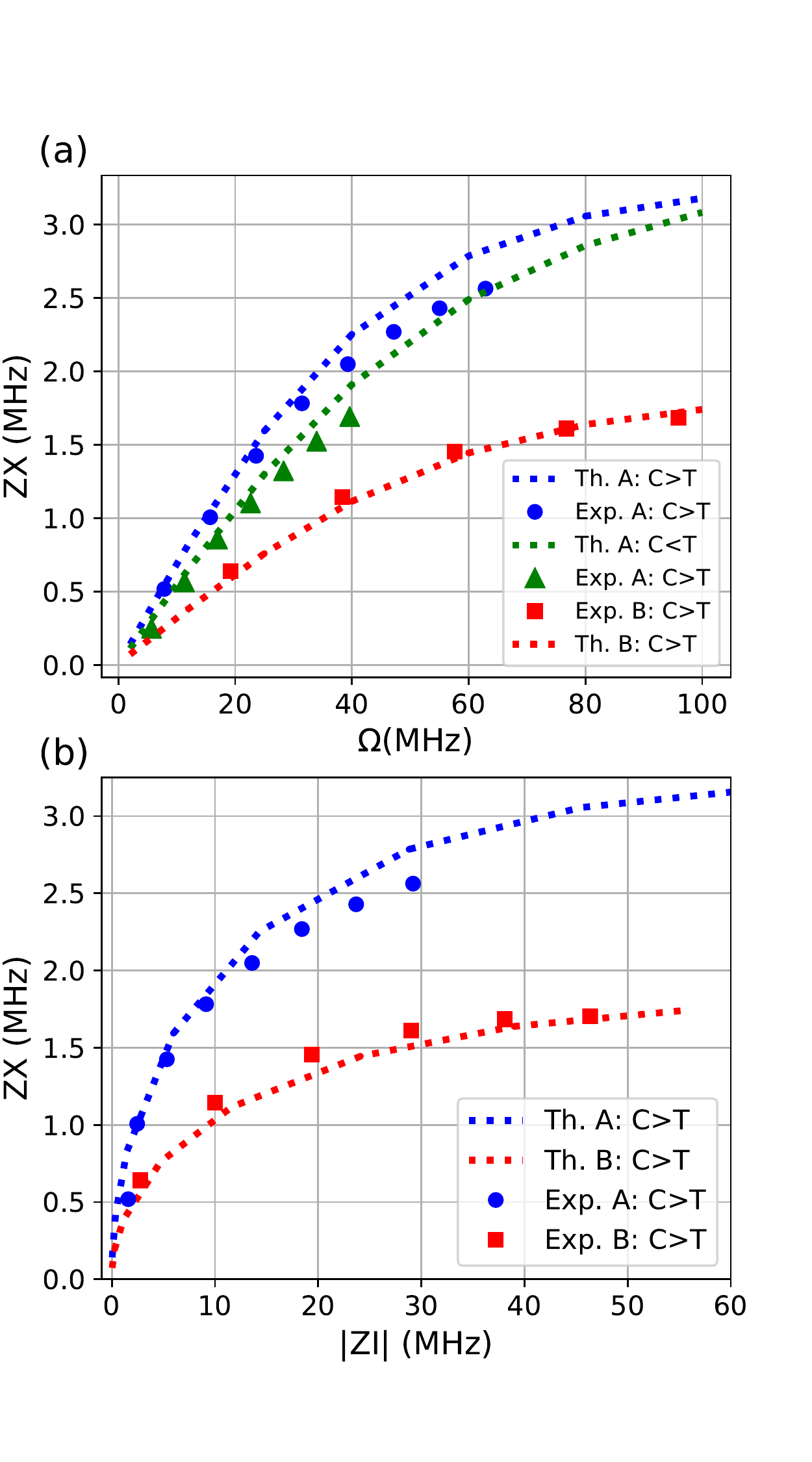}
\caption{(a) ZX vs CR drive strength ($\Omega$) calculated from theory (dashed lines) for device A and for a single-path coupler device B. Data points show the data measured experimentally. Although the single coupling device has higher ZZ, it has a lower $ZX$ rate. For device A we measure in both CR directions; when the control (C) is higher frequency than the target (T) and vice-versa. (b) The magnitude of the control stark shift ($|ZI|$) versus the $ZX$ rate for the two devices (the shift is negative). At the same Stark shift, device A supports much larger $ZX$ rates, attributed to the greater $J_\textrm{eff}$. \label{fig:zx}}
\end{figure}

We now delve into the dynamic properties of the device under CR drives, which entails driving a control qubit at the frequency of the target qubit, with an amplitude $\Omega$. While the desired entangling interaction is $ZX$, the drive Hamiltonian constitutes several unwanted terms that have been studied extensively in theory and experiment~\cite{sheldon:2016,sundaresan:2020,Magesan2020}. This includes a control qubit stark shift $ZI$ that is a consequence of the off-resonant tone on the control qubit. While the $ZI$ interaction is often nullified by echo-sequences, the additional single qubit gates and pulse ramps lead to a gate time cost. Instead, the approach we introduce here involves the use of calibrated frame-changes~\cite{mckay:2017} on the control qubit to null the Stark shift, which has no additional time cost. However, this relies on the stability of the Stark shift, which is intrinsically related to amplitude noise of the CR pulse $ZI \sim \Omega^2$~\cite{magesan:2020}. In Fig.~\ref{fig:zx}, we study the $ZX$ and $ZI$ interaction rates as a function of drive amplitude, measured using Hamiltonian tomography~\cite{sheldon:2016} and Ramsey sequences, respectively. The experimental data shows good agreement with numerical simulations. Note that the low drive, linear $ZX$ limit in Fig.~\ref{fig:zx}(a) is employed to estimate the $J_{\textrm{eff}}$ discussed previously. We also compare these interaction rates to device B, which has qubit frequencies $f_0 (f_1)$ = 5.1330 (5.0442) GHz,  $\alpha_0 (\alpha_1)$ = -318 (-320) MHz and average coherence $T_1=[101(2),121(8)]~\mu s$ and $T_2=[90(4),100(4)]~\mu s$ (similar to device A). Device B has a single direct capacitive coupler with $J=2.07$~MHz corresponding to a $ZZ=58$~kHz; larger than device A despite the lower $J$ due to the lack of a multi-path coupler's $ZZ$ cancellation. The effect of enhanced $J_\textrm{eff}$ is apparent in the comparatively larger $ZX$ rates for device A, enabling faster two qubit gates. Furthermore, this also translates into a comparatively smaller $ZI$ on device A for a desired $ZX$ rate, seen in Fig.~\ref{fig:zx}(b). This can have important implications for the stability of un-echoed two-qubit gates constructed from CR pulses, as will be detailed next.

We finally discuss the construction of a CNOT gate with our device architecture and cross-resonance. The CNOT gate is particularly useful for many algorithms, and is also advantageous for benchmarking, since it belongs to the Clifford group. Typical CNOT constructions with CR have employed echo sequences~\cite{corcoles:2013,sheldon:2016} sandwiched between single qubit rotations. An alternate approach uses only a single CR pulse and single qubit operations, dubbed a direct CNOT, that is more efficient in total gate-time but is not naturally insensitive to low frequency amplitude noise (a similar direct CNOT was also used recently in Ref~\cite{jurcevic:2020}). The direct CNOT gate is constructed from two physical pulses that are applied simultaneously: a CR drive on the control qubit, and a resonant drive on the target qubit. Following a rough amplitude calibration of the CR pulse for a chosen gate time, the phase of the CR drive is calibrated to minimize the $ZY$ term in Hamiltonian tomography ~\cite{sheldon:2016}, with both calibrations performed in the absence of a target drive. This is followed by a simultaneous fine calibration (using error amplification sequences~\cite{sheldon:2015}) of the CR/target drive amplitude, target DRAG, and CR/target drive phases such that the resultant target dynamics is a $2\pi$ rotation when the control is in $\ket{0}$ and a $X_\pi$ rotation when the control is in $\ket{1}$. The gate unitary now can be written as $U=\ketbra{0}\otimes I + e^{i \phi} \ketbra{1}\otimes X$, where $\phi$ is a phase on the control qubit generated by the CR drive, related to its Stark shift. Finally, we add a frame change~\cite{mckay:2017} on the control qubit at the end of the gate to cancel $\phi$, which brings the unitary to the desired CNOT gate. As discussed previously, the suppressed Stark shift in device A plays an important role in the stability of this frame change.

\begin{figure}
\includegraphics[width = \columnwidth]{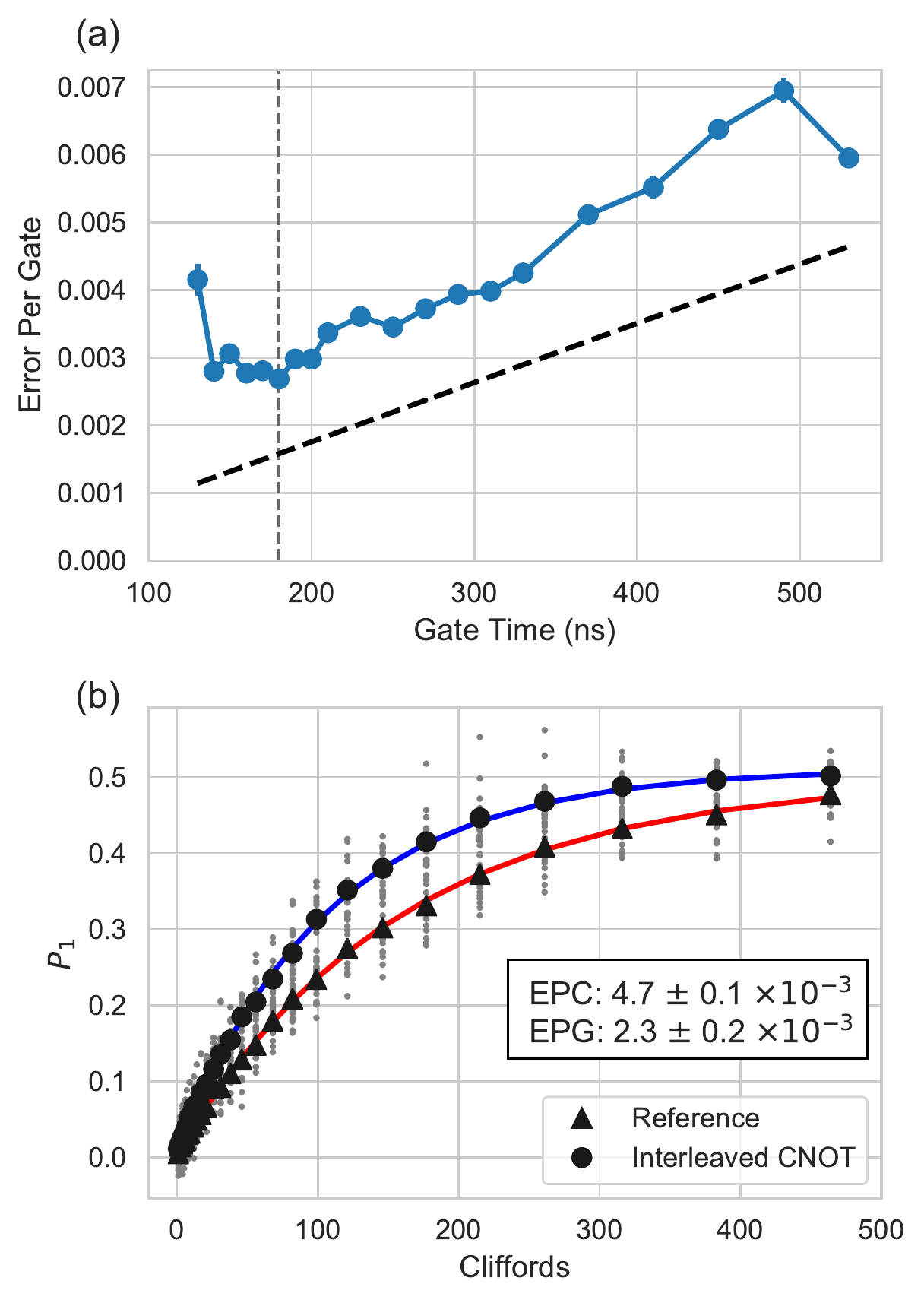}
\caption{(a) Finding the gate-time that optimizes the error. At each point we perform standard RB, measure the error per Clifford (EPC) and divide by the number of CNOT gates per Clifford (see supplement) to arrive at the error per gate (EPG). This is an upper bound on the EPG since it assumes the single qubit error contribution to the EPC is zero. The dashed line is the estimated lower bound EPG based on the measured $T_1,T_2$.  (b) At the optimal gate length of 180~ns, vertical dashed line in (a), we perform interleaved RB. Averaging over the measurements on the two qubits, the EPG is 2.3$\times~10^{-3}$ (fidelity of 99.77\%) and the EPC is 4.67$\times 10^{-3}$ which gives an error upper bound of 3.0$\times 10^{-3}$. \label{fig:gate}}
\end{figure}

In Fig.~\ref{fig:gate} we show the results of our gate optimization for various gate-times. Fig.~\ref{fig:gate}(a) reports an upper bound on the gate error (see caption) as a function of the average gate-time, with a characteristic upturn at the shortest times. At the optimal length of 180~ns, we show interleaved randomized benchmarking~\cite{magesan:2012} curves in Fig.~\ref{fig:gate}(b), that we use to estimate a two qubit gate error of only 2.3$\times 10^{-3}$ (upper bound of 3.0$\times 10^{-3}$ from standard RB). Additional characterization of the gate reveals that the measured error rate is consistent with purity benchmarking, and leakage contributions to the error to be less than $10^{-4}$ (see supplement). It is important to highlight that the enhanced $J_{\textrm{eff}}$ and suppressed static $ZZ$, enable both: a state-of-the art CNOT gate constructed using cross-resonance, and the high-fidelity, simultaneous operation of 40~ns single qubit gates at an error of 3.5(1)$\times 10^{-4}$ and 2.7(1)$\times 10^{-4}$ for Q0 and Q1 respectively. This manifests in the reference RB decay of Fig.~\ref{fig:gate}(b) extending to $\sim 500$ two-qubit Clifford operations.

In conclusion, we demonstrate a fixed frequency architecture for transmons with an engineered suppression of the $ZZ$ interaction term through the use of two elements -- a direct capacitive coupler and an $\lambda/4$ resonator. This multi-path coupler allows the increase of effective $J$ coupling between the qubits, without the corresponding unwanted $ZZ$ interaction, i.e., breaking the standard $J/ZZ$ relationship of single element couplers. This enables us to realize a single pulse CNOT with an error of 2.3$\times 10^{-3}$, more than a factor of two improvement over the previous best reported fidelity of 5$\times 10^{-3}$~\cite{jurcevic:2020} for a cross-resonance CNOT gate. Since fixed-frequency superconducting processors with over 60 qubits have already been demonstrated based on cross-resonance, this work provides a clear path for superior multi-qubit circuit performance via faster two qubit gates and reduced $ZZ$ error, without any degradation of coherence or increase in control complexity. \\

\begin{acknowledgments}
We thank Muir Kumph, Shawn Hall and Vincent Arena for help with packaging. We acknowledge helpful discussions with Isaac Lauer, Neereja Sundaresan, Firat Solgun, Pranav Mundada, Gengyan Zhang, Thomas Hazard and Andrew Houck. The gate bring-up, investigation, and characterization work was supported by IARPA under LogiQ (contract W911NF-16-0114).
\end{acknowledgments}

\bibliography{maple}

\pagebreak
\widetext
\begin{center}
\textbf{\large Supplemental Materials: Demonstration of a High-Fidelity CNOT for Fixed-Frequency Transmons with Engineered ZZ Suppression}
\end{center}
\setcounter{equation}{0}
\setcounter{figure}{0}
\setcounter{table}{0}
\setcounter{page}{1}
\makeatletter
\renewcommand{\theequation}{S\arabic{equation}}
\renewcommand{\thefigure}{S\arabic{figure}}
\renewcommand{\bibnumfmt}[1]{[S#1]}
\renewcommand{\citenumfont}[1]{S#1}

\section{Device Modeling}

To measure the coupler-mode frequencies and coupling strengths, $\omega_j$ and $g_{i,j}$ from Eqn.~\ref{eqn:mainh}, we used a number splitting technique~\cite{schuster:2007}. Here we calibrated a very slow ($\approx \mu$s) $\pi$ pulse on the qubit and applied it  after driving the qubit at the cavity frequency. Since any photon population in the cavity results in a $\chi$ shift to the qubit on the few MHZ level, the $\pi$ pulse will not complete a full rotation when photons are in the cavity. Therefore we measure the cavity frequency by sweeping the drive and observing the drive frequency for which the qubit population in $|1\rangle$ after the $\pi$ pulse is reduced . Once the cavity frequency is found, we can look at the precise single photon $\chi$ shift on the qubit from which $g$ can be inferred.  \\

To solve for $ZZ$ and $\mu$ we diagonalize Eqn.~\ref{eqn:mainh}. We calculate $ZZ$ as $\tilde{E}_{11}+\tilde{E}_{00}-(\tilde{E}_{01}+\tilde{E}_{10})$ where $\tilde{E}_{ij}$ is the dressed energy of the $i$ state of dressed transmon 0 and $j$ state of dressed transmon 1 (defined as the states which adiabatically follow from the bare transmon states). $\mu$ is defined as the $ZX$ rate per applied Rabi drive and is calculated as $\tilde{\langle 11|} \hat{a}^{\dagger}_0 \tilde{|10\rangle} - \tilde{\langle 01|} \hat{a}^{\dagger}_0 \tilde{|00\rangle}$ for the case in which qubit 0 is the control and $\hat{a}^{\dagger}_0$ is the bare raising operator for transmon 0 and $\tilde{|ij\rangle}$ is the dressed eigenstate. \\

The numerical $ZX$ rate is calculated by diagonalizing Eqn.~\ref{eqn:mainh}, then applying a rotating frame at the target frequency $H = H_0-\omega_{T}(\tilde{\hat{n}}_0+\tilde{\hat{n}}_1) + \Omega \sum_{ij} \tilde{|i\rangle} \tilde{\langle i|} \hat{a}_0^{\dagger} \tilde{|j\rangle} \tilde{\langle j|}$. This Hamiltonian is truncated to 3 levels per transmon and solved numerically versus pulse time with the control in $|0\rangle$ or $|1\rangle$ to extract the $ZX$ rate.  

\section{Parameter Sensitivity}

\begin{figure}
\includegraphics[width = 0.9\textwidth]{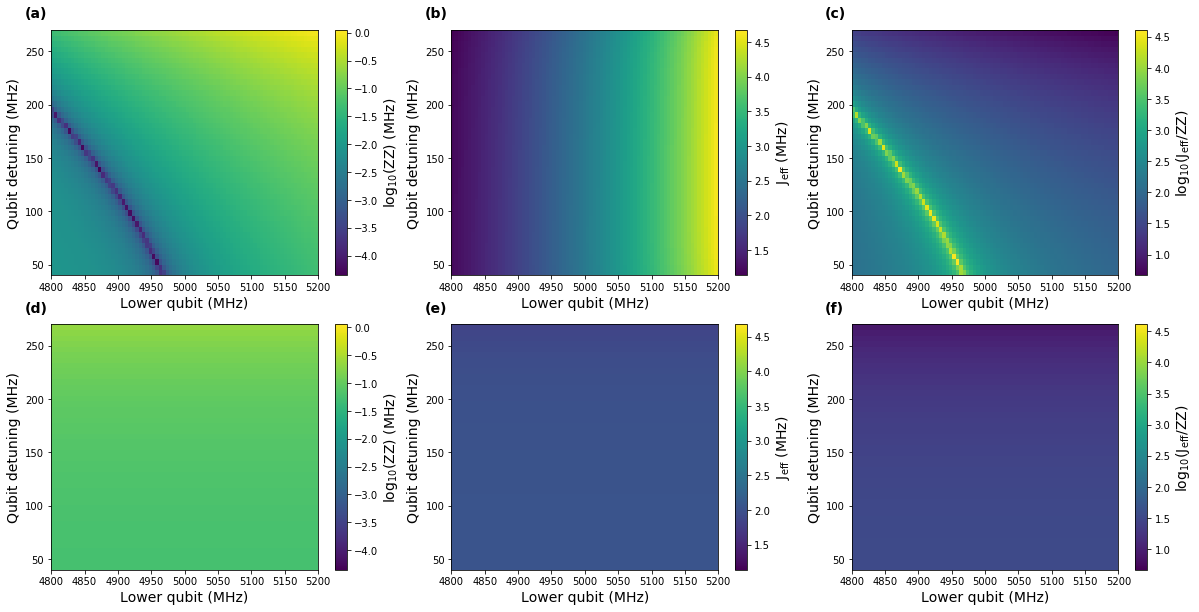}
\caption{$ZZ$, $J_{\textrm{eff}}$, and $J_{\textrm{eff}}/ZZ$ swept for a range of qubit frequencies with the coupling parameters of device A ((a), (b), (c)) and device B ((d), (e), (f)). \label{fig:sensitivity}}
\end{figure}

Although our fixed frequency architectures are very favorable for accessing state-of-the-art coherence times and ease of quantum control, uncertainties in device fabrication can lead to frequency variability. This is particularly a challenge for devices with single junction transmons devices, where frequency collisions can lead to gate infidelity~\cite{magesan:2020,Hertzberg:2020}. In our coupling scheme there is an additional constraint since the region of $ZZ$ cancellation is dependent on the qubit frequencies, as seen in Fig.~\ref{fig:schematic} of the main text. Further, the efficacy of the $ZZ$ cancellation can potentially be exacerbated by uncertainties in the frequencies of {\it{both}} qubits. Therefore, we numerically study the performance of our cancellation couplers for a wide range of qubit frequencies ($>$ 200 MHz) in the straddling regime, with all other Hamiltonian parameters kept fixed. In Fig.~\ref{fig:sensitivity}, we employ the coupling parameters of device A and B, introduced in the main text, and study the variability of $ZZ$, $J_{\textrm{eff}}$, and $J_{\textrm{eff}}/ZZ$ over this large spread in qubit frequencies. While the region of $ZZ \sim 0$ remains very sensitive to the frequency placement, we show that almost over the entire range of frequencies considered, the cancellation coupler outperforms our standard coupler in both, $ZZ$ as well as $J_{\textrm{eff}}/ZZ$. We also note that higher $J_{\textrm{eff}}$ are accessible with the cancellation couplers away from $ZZ = 0$ operating points -- this is of relevance when optimizing for finite device coherence.

\section{Pulse Shapes}

\begin{figure}
\includegraphics[width = 0.5\textwidth]{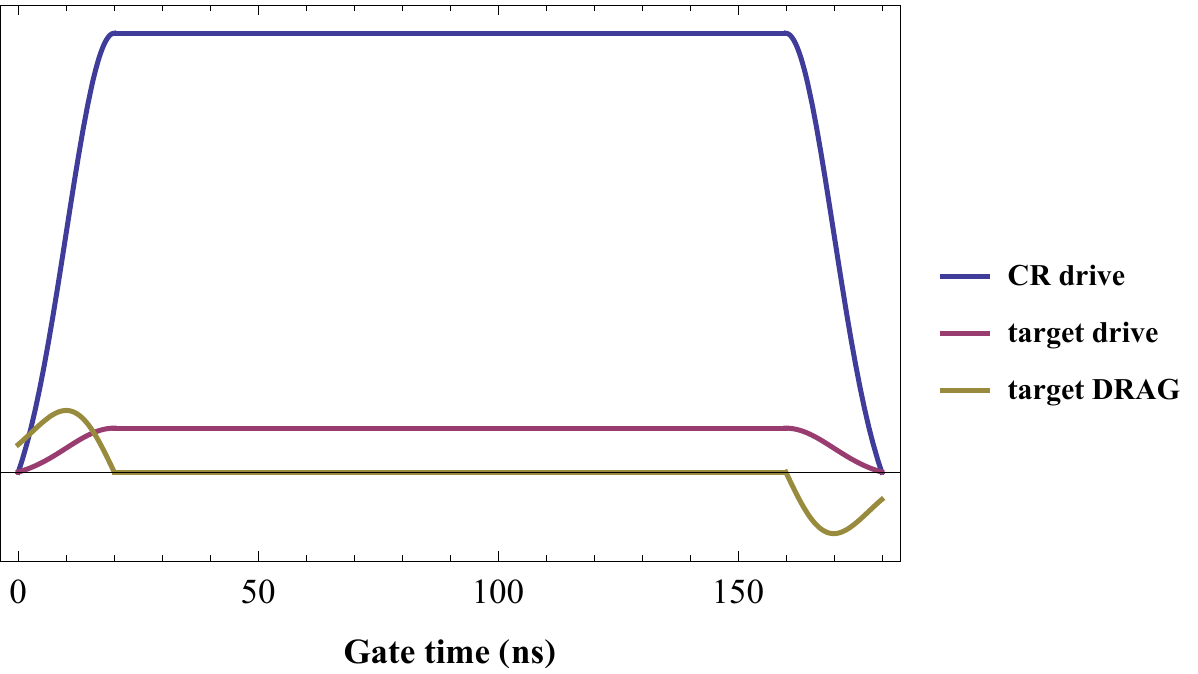}
\caption{The pulse shapes for the cross-resonance (CR) drive, target drive, and target DRAG used in constructing the direct CNOT gate. The target DRAG is applied 90 degrees out of phase with respect to the target drive. The vertical dimension is not drawn to scale. \label{fig:pulseshape}}
\end{figure}

The pulse shape used for single qubit gates is a Gaussian envelope with $\sigma=10 $ns truncated at $\pm 2\sigma$ and lowered (to start at zero amplitude) with derivative Gaussian quadrature correction (DRAG). The pulse used for the CNOT gate consists of two simultaneous components: a cross-resonance drive applied on the control qubit at the target qubit's frequency and an on-resonance drive applied to the target qubit. The pulse shape for both drives is a flat-topped Gaussian, with rise and fall time of $2\sigma$ and $\sigma=10 $ns. The on-resonance target pulse has derivative Gaussian quadrature correction. The pulse shape for the CNOT gate is shown in Fig.~\ref{fig:pulseshape}; the vertical dimension is not drawn to scale.

\section{Clifford Gateset}

For our 2Q Clifford gateset we use a combination of finite-length $\pi/2$ pulses along the x and y axes, software $Z$ gates~\cite{mckay:2017} which do not add any additional length to the Clifford, and the CNOT gate. We write the single-qubit gates as ``rotation axis''-``rotation angle(in degrees)''-``(p)ositive/(m)inus''. For example, X90m, is a $-\pi/2$ rotation around the x axis. The average number of each gate per Clifford is as follows:

\begin{center}
\begin{tabular}{|c|c|}
\hline
Gate & Average number per Clifford \\ \hline \hline
X90m & 0.4477 \\ \hline
X90p & 0.6686 \\ \hline
Z90m & 0.5302 \\ \hline
Z0 & 0.2951 \\ \hline
Z90p & 1.1149 \\ \hline
Y90p & 0.9467 \\ \hline
Zp & 0.9465 \\ \hline
Y90m & 0.5981 \\ \hline
CNOT & 1.5712 \\ \hline
\end{tabular}
\end{center}

Therefore, there are 2.66 non-Z single-qubit gates per Clifford. For the data shown in Fig.~\ref{fig:gate} (b), the average Clifford gate is 389~ns long. 

\section{Further Gate Characterization}

\begin{figure}
\includegraphics[width = 0.75\textwidth]{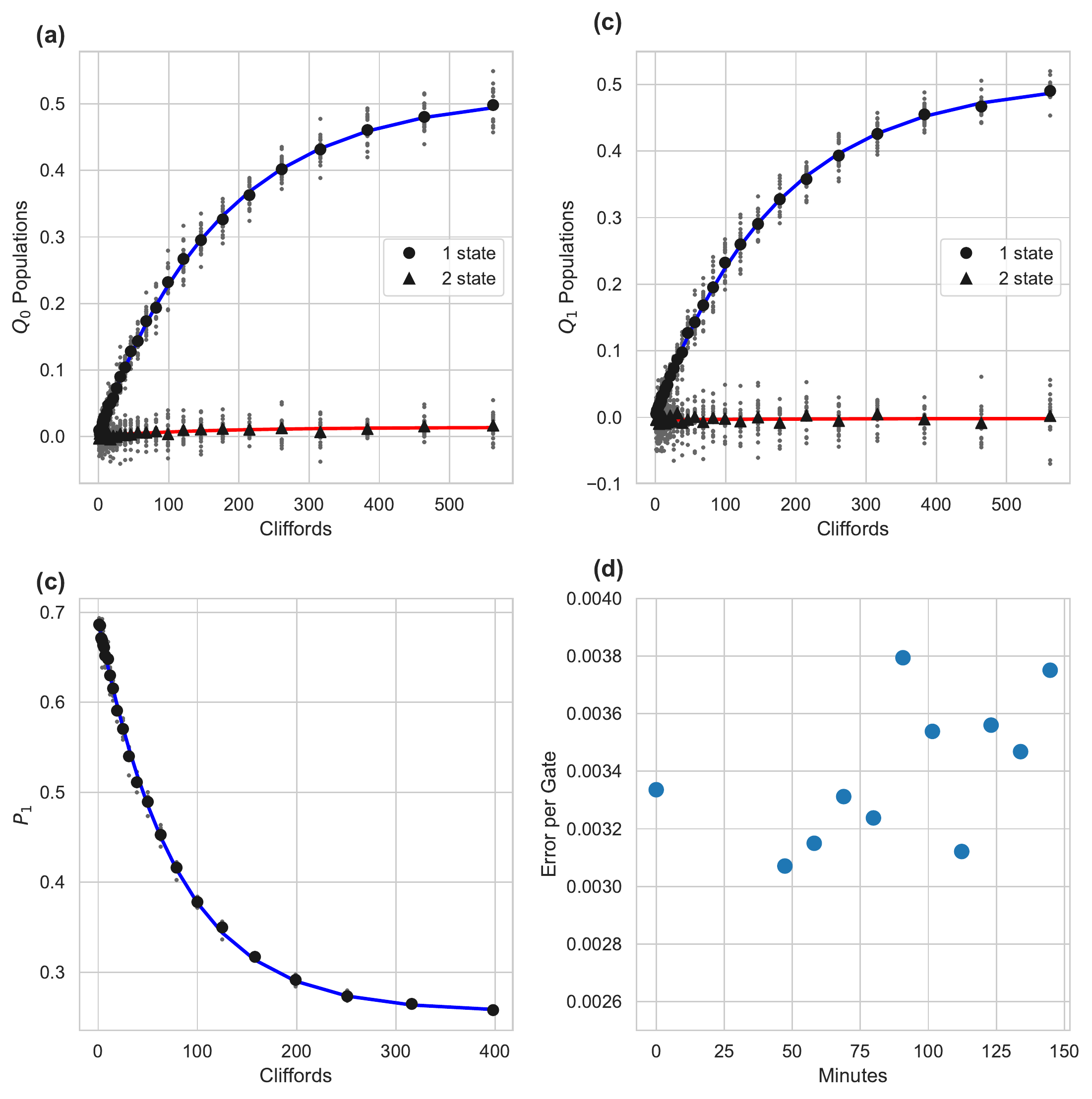}
\caption{(a),(b) leakage randomized benchmarking and (c)  purity randomized benchmarking. (d) Stability of the gate over a period of approximately two hours without recalibration. The gate length here is 280~ns and the error per gate is an upper bound extracted from standard randomized benchmarking.\label{fig:gate_more}}
\end{figure}

We show a suite of further gate characterizations in Fig.~\ref{fig:gate_more}. In (a) and (b) we measure leakage RB~\cite{mckay:2017} on both qubits, i.e., we measure both the $|1\rangle$ and $|2\rangle$ state population during a standard RB sequence. By fitting the $|2\rangle$ state population to an exponential decay, we extract leakage rates per Clifford of 9(1)$\times10^{-5}$ and -1(3)$\times10^{-5}$ for Q0 and Q1. Unsurprisingly the leakage on the target is effectively zero. The leakage on the control is measurable, but still an order of magnitude lower than the gate error. After more than 500 Cliffords the control population in $|2\rangle$ is about 1\%. In (c) we show the results of purity RB~\cite{mckay:2016}, where we perform standard RB sequences, but measure the trace of $\rho^2$ by combining the results of several measurements with appropriate post rotations (the data is not corrected for readout errors).  We fit to a depolarizing error model and measure an effective depolarizing error per Clifford of 4.78(3)$\times 10^{-3}$, which is consistent with what we measured from standard RB. This indicates the gate is primarily limited by incoherent errors, as opposed to calibration errors or coherent crosstalk such as $ZZ$. Finally, in (d) we show the stability of the gate calibration over a period of approximately two hours. 

\end{document}